\documentclass[review,number,sort&compress]{elsarticle}
\usepackage{lineno}
\usepackage{url}
\usepackage{amssymb}
\usepackage{amsmath}
\usepackage{mathtools}
\usepackage{multirow}
\usepackage{array}
\usepackage{hyperref}
\usepackage{subcaption}
\usepackage{comment}
\usepackage{makecell}
\usepackage{float}
\usepackage{makecell}
\usepackage{adjustbox}
\usepackage{xcolor}
\usepackage{ulem}
\modulolinenumbers[5]

\begin{document}

\begin{frontmatter}

\title{A Novel Low-Background Photomultiplier Tube Developed for Xenon Based Detectors}

\author{Youhui Yun \corref{cor1}\fnref{fn0}}
\author{Zhizhen Zhou \corref{cor1}\fnref{fn0}}
\author{Baoguo An\fnref{fn8}}
\author{Zhixing Gao\fnref{fn0}}
\author{Ke Han\fnref{fn0,fn1,fn2}}
\author{Jianglai Liu\fnref{fn0,fn1,fn2,fn3}}
\author{Yuanzi Liang\fnref{fn4}}
\author{Yang Liu\fnref{fn0}}
\author{Yue Meng \corref{cor2}\fnref{fn0,fn1,fn2}}
\author{Zhicheng Qian\fnref{fn0}}
\author{Xiaofeng Shang\fnref{fn0}}
\author{Lin Si\fnref{fn0}}
\author{Ziyan Song\fnref{fn8}}
\author{Hao Wang\fnref{fn0}}
\author{Mingxin Wang\fnref{fn0}}
\author{Shaobo Wang\fnref{fn0,fn1,fn2}}
\author{Liangyu Wu\fnref{fn0}}
\author{Weihao Wu\fnref{fn0}}
\author{Yuan Wu\fnref{fn0}}
\author{Binbin Yan \corref{cor2}\fnref{fn5}}
\author{Xiyu Yan\fnref{fn6}}
\author{Zhe Yuan\fnref{fn7}}
\author{Tao Zhang\fnref{fn5}}
\author{Qiang Zhao\fnref{fn8}}
\author{Xinning Zeng\fnref{fn0}}

\address[fn0]{Shanghai Key Laboratory for Particle Physics and Cosmology, Institute of Nuclear and Particle Physics (INPAC) and School of Physics and Astronomy, Shanghai Jiao Tong University, Shanghai 200240, China}
\address[fn8]{Hamamatsu Photonics (China) Co.,Ltd}
\address[fn1]{Sichuan Research Institute, Shanghai Jiao Tong University, Chengdu 610213, China}
\address[fn2]{Jinping Deep Underground Frontier Science and Dark Matter Key Laboratory of Sichuan Province}
\address[fn3]{New Cornerstone Science Laboratory, Tsung-Dao Lee Institute, Shanghai Jiao Tong University, Shanghai 201210, China}
\address[fn4]{College of Physics, Jilin University, Jilin 130012, China}
\address[fn5]{Tsung-Dao Lee Institute, Shanghai Jiao Tong University, Shanghai 201210, China}
\address[fn6]{School of Physics and Astronomy, Sun Yat-Sen University, Zhuhai 519082, China}
\address[fn7]{Key Laboratory of Nuclear Physics and Ion-beam Application (MOE), Institute of Modern Physics, Fudan University, Shanghai 200433, China}

\cortext[cor1]{Co-first authors.}
\cortext[cor2]{Corresponding authors: mengyue@sjtu.edu.cn and yanbinbin@sjtu.edu.cn}

\begin{abstract}

Photomultiplier tubes (PMTs) are essential in xenon detectors like PandaX, LZ, and XENON experiments for dark matter searches and neutrino properties measurement. To minimize PMT-induced backgrounds, stringent requirements on PMT radioactivity are crucial. A novel 2-inch low-background R12699 PMT has been developed through a collaboration between the PandaX team and Hamamatsu Photonics K.K. corporation. Radioactivity measurements conducted with a high-purity germanium detector show levels of approximately 0.08 mBq/PMT for $\rm^{60}Co$ and 0.06~mBq/PMT for the $\rm^{238}U$ late chain, achieving a 15-fold reduction compared to R11410 PMT used in PandaX-4T. The radon emanation rate is below 3.2 $\rm \mu$Bq/PMT (@90\% confidence level), while the surface $\rm^{210}Po$ activity is less than 18.4 $\mu$Bq/cm$^2$. The electrical performance of these PMTs at cryogenic temperature was evaluated. With an optimized readout base, the gain was enhanced by 30\%, achieving an average gain of $4.23 \times 10^6$ at -1000~V and -100~$^{\circ}$C. The dark count rate averaged 2.5~Hz per channel. Compactness, low radioactivity, and robust electrical performance in the cryogenic temperature make the R12699 PMT ideal for next-generation liquid xenon detectors and other rare event searches.

\end{abstract}

\end{frontmatter}

\section{Introduction}
\label{sec:intro_merge}

\begin{figure}[htbp]
    \centering
    \includegraphics[width=.7\textwidth]{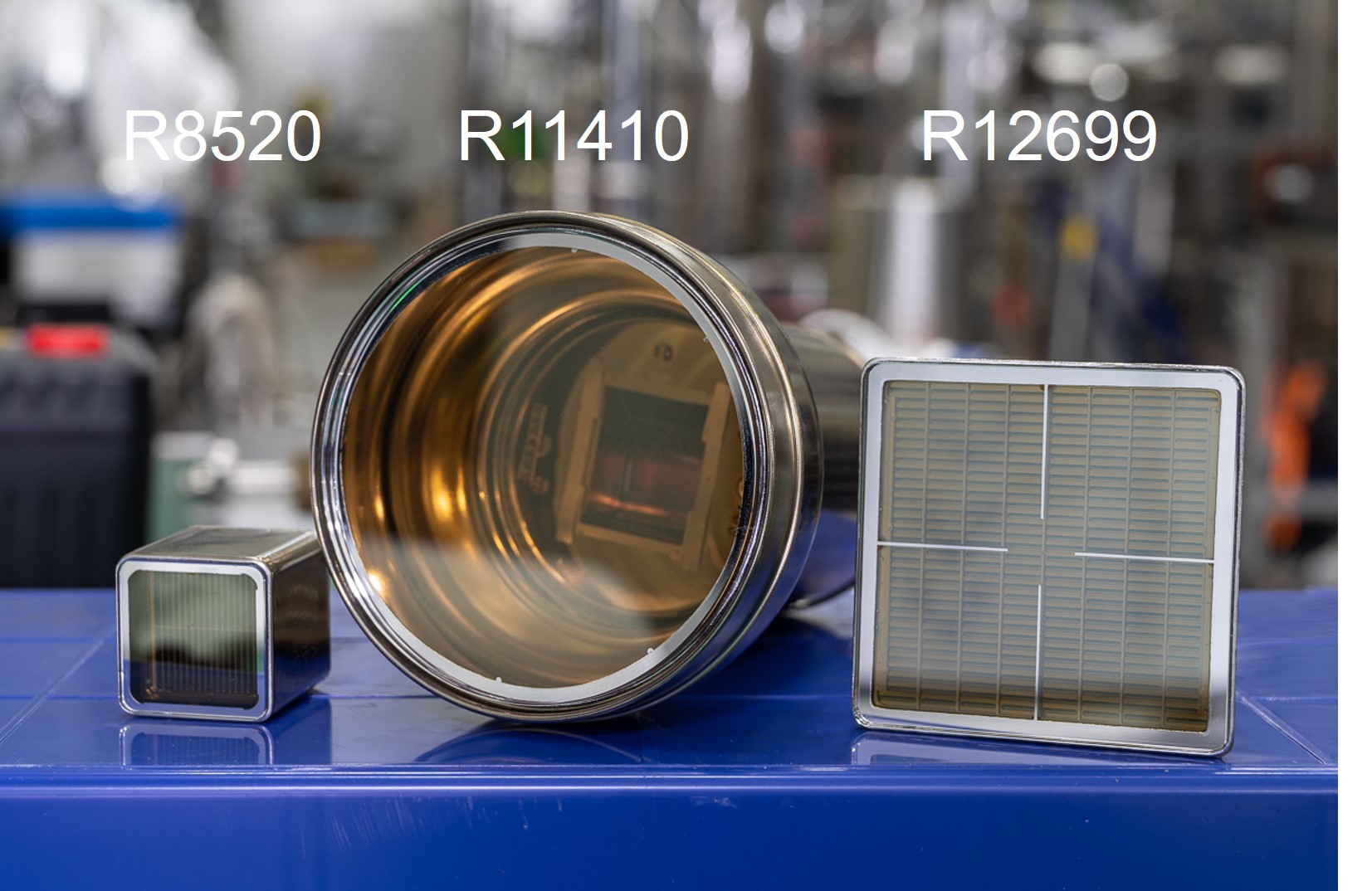}
    \caption{The three types of PMTs shown here are applied for the three generations of liquid xenon detectors. The R8520 PMT on the left for the first generation, the R11410 PMT in the middle is currently used, and the R12699 PMT on the right is a potential candidate for the next generation liquid detectors. The R12699 PMT integrates four independent detecting channels into one tube.}
    \label{fig:pmt}
\end{figure}

Liquid xenon detectors are commonly used in the search for dark matter (DM) and the study of neutrino properties. The search for dark matter focusing on the weakly interacting massive particles (WIMPs), a hypothesized class of particles proposed to explain dark matter \cite{k}, has recently been constrained by the stringent limit on the spin-independent WIMP-nucleon cross section, which is set at 1.6 $\times 10^{-47}$cm$^2$ for 40 GeV/c$^2$~\cite{WIMP_run01}. For neutrino, the studies focus on the neutrinoless double beta decay (NLDBD), a rare nuclear process that would confirm neutrinos as Majorana particles~\cite{Furry:1939qr}. These liquid xenon detectors utilize a dual-phase xenon time projection chamber (TPC) technique. When incident particles interact with xenon atoms in liquid xenon (LXe), they deposit energy, causing the excitation and ionization of the xenon atoms. Some of the ionized electrons re-combine with xenon atoms, emitting scintillation light with a wavelength centered around 178~nm \cite{a,l}, known as the primary light signal ($S1$). The remaining ionized electrons are drifted to the surface of LXe by an electric field and extracted into the gaseous xenon (GXe) to produce a secondary scintillation signal ($S2$) by electroluminescence \cite{v}. The paired $S1$ and $S2$ signals can be detected by effective vacuum ultraviolet (VUV) photon detectors, providing comprehensive information about each event. 1-inch Hamamatsu R8520 photomultiplier tubes (PMTs)~\cite{R8520}, shown on the left of Figure\ref{fig:pmt}, have been used in the PandaX-I~\cite{PandaX-I} and XENON10~\cite{c} experiments. 3-inch Hamamatsu R11410 PMTs \cite{d}, shown in the middle of Fig.~\ref{fig:pmt}, have been widely used as the primary photon detector in various tonne-scale liquid xenon experiments, such as XENON1T \cite{p}, XENONnT \cite{XENONnT}, LUX-ZEPLIN \cite{n,u}, and PandaX-4T \cite{o}.

Next-generation proposed xenon experiments, such as XLZD~\cite{XLZD}, DARWIN~\cite{DARWIN}, KamLAND2-Zen~\cite{Nakamura:2020szx}, and PandaX-xT~\cite{PandaX-30T}, which feature multi-ten-tonne active targets, aim to simultaneously search for both WIMPs and NLDBD events. These experiments seek to enhance the sensitivity for detecting WIMPs, approaching the neutrino floor \cite{r}, thereby providing a decisive verdict on the existence of WIMPs and enabling the measurement of neutrino properties~\cite{q}. To meet the stringent background requirements of these next-generation detection experiments, a new type of PMT, the R12699-406-M4 (as shown on the right of Fig.~\ref{fig:pmt}, R12699 for short), has been developed in cooperation with Hamamatsu Photonics K.K corporation. The geometric sizes of R12699, R11410, and R8520 are shown in Tab.~\ref{tab:PMT_size}.

\begin{table}[htbp]
    \caption{The size of R8520, R11410 and R12699 PMTs. Unit:~mm.}
    \label{tab:PMT_size}
    \centering
    \resizebox{\textwidth}{!}{
    \begin{tabular}{c|c|c|c}
    \hline
    PMT model&Outer diameter/side length&Cathode diameter/side length& Height (excluding pins)\\
    \hline
    R8520 &30&20.5&28.25\\
    \hline
    R11410&76&64&114\\
    \hline
    R12699&56&48.5&14.8\\
    \hline
    \end{tabular}
    }
\end{table}

\begin{figure}[!htbp]
\centering
    \begin{minipage}[t]{0.6\linewidth}
        \centering
        \includegraphics[width=\linewidth]{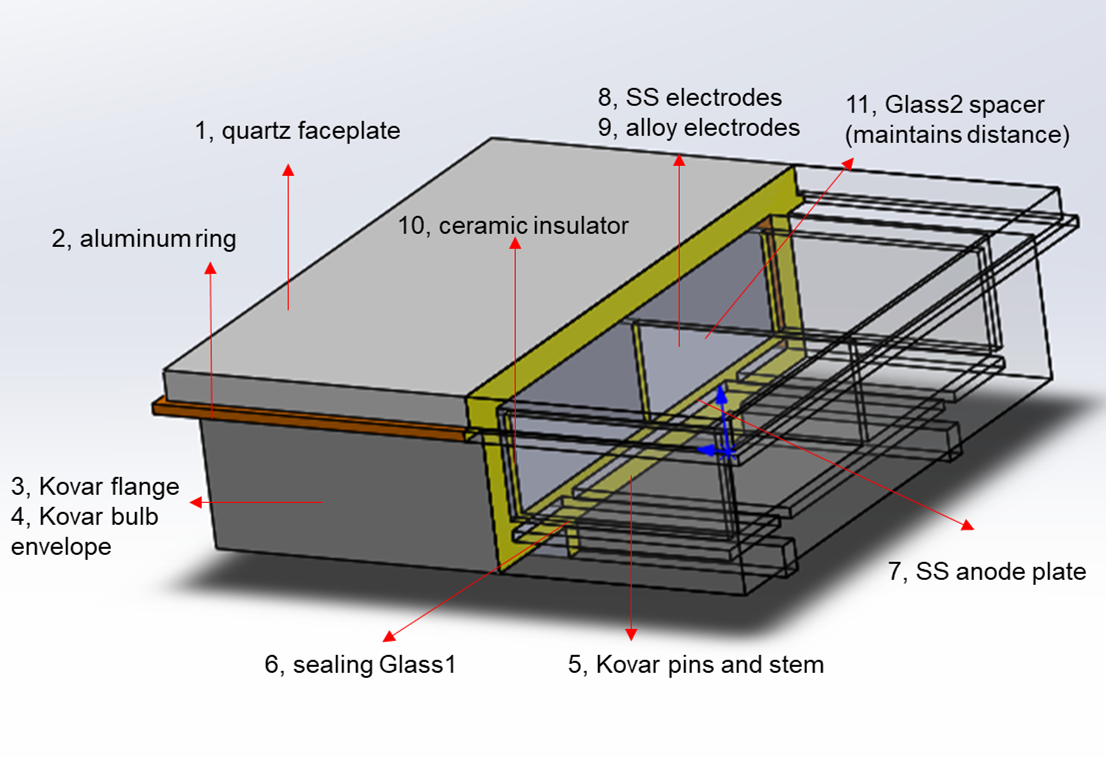}
        \caption{The structure of the R12699 PMT.}
        \label{figPMTStructure}
    \end{minipage}
\end{figure}

The R12699 PMT is a 2-inch square, compact quartz-windowed PMT with a bialkali cathode material (the same cathode material used in the R11410 PMT), sensitive to wavelengths from 160 nm to 650 nm, and designed to operate down to -110~$^{\circ}$C \cite{e}. This model features four identical detecting channels integrated into one tube, making it suitable for a higher-density layout of photo-detecting surface. In addition, it exhibits fast time response performance due to the compact dynode construction, with a rise time of approximately 1.2 ns, a transit time of 5.9 ns, and a transit time spread of 0.41 ns. As illustrated in Fig.~\ref{figPMTStructure}, the structure of the R12699 PMT comprises a quartz faceplate window that facilitates the penetration of VUV light, electrodes, ceramic insulators, and stainless steel (SS) anodes within a metal body. The different parts are numbered for clarity in the subsequent discussion.

In this article, we describe the background improvements of R12699 PMTs, including radioactivities of the bulk material, radon emanation rate and surface radioactivity. We also outline the cryogenic test setup and the electrical performance of a batch of 54 R12699 PMTs at cryogenic temperature. Characteristic parameters such as gain, dark count rate (DCR), and after-pulse (AP) probability are measured. The screening methods and measurement results of the background are detailed in Section~\ref{screen}. The cryogenic testing setup is detailed in Section~\ref{sec:setups}. Section~\ref{sec:result} presents the test results of PMT electrical performances. The conclusions and discussions are provided in Section~\ref{sec:sum}, where we also compare the characteristics of R11410 and R12699 PMTs.

\section{Screening methods and results of background}
\label{screen}
In the background control of rare event search experiments, our primary focus is monitoring radioactive isotopes and decay chains.
Besides, the radon emanation and surface $^{210}$Po activity of PMT also induce backgrounds. This section describes low background assay techniques including the High Purity Germanium (HPGe) detector, radon emanation detector, and alpha detector. The measurement results and radioactivity improvement are also shown here.

\subsection{Radioactivities of bulk material}
\label{RI_bulk}
The bulk materials contain radioactive isotopes, including $\rm^{60}Co$, $\rm^{40}K$, $\rm^{137}Cs$, as well as decay chains  such as $\rm^{235}U$, $\rm^{232}Th$, and $\rm^{238}U$.
The decay chain of $\rm^{238}U$ is divided into early and late chains due to the disruption of secular equilibrium at $\rm^{222}Rn$. Similarly, in the $\rm^{232}Th$ chain, this division occurs at $\rm^{228}Th$.
The $\rm^{238}U$ late chain ($\rm^{238}U(l)$), the $\rm^{232}Th$ late chain ($\rm^{232}Th(l)$) and $\rm^{60}Co$ are particularly important as they will induce backgrounds in the region of interest of DM searching and NLDBD. 
The radioactivity of these isotopes can be measured through their gamma emissions.

\begin{figure}[!htbp]
    \centering
    \includegraphics[width=0.45\textwidth]{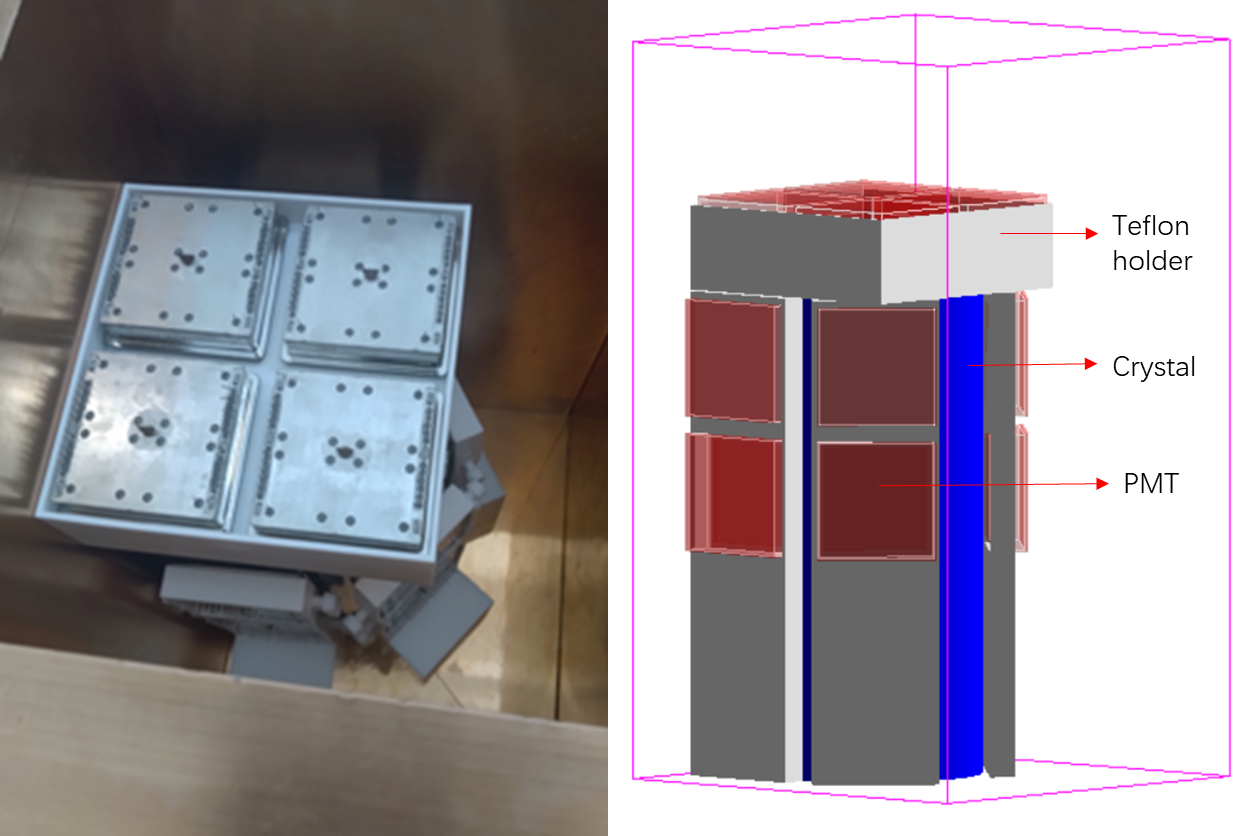}
    \qquad
    \includegraphics[width=0.47\textwidth]{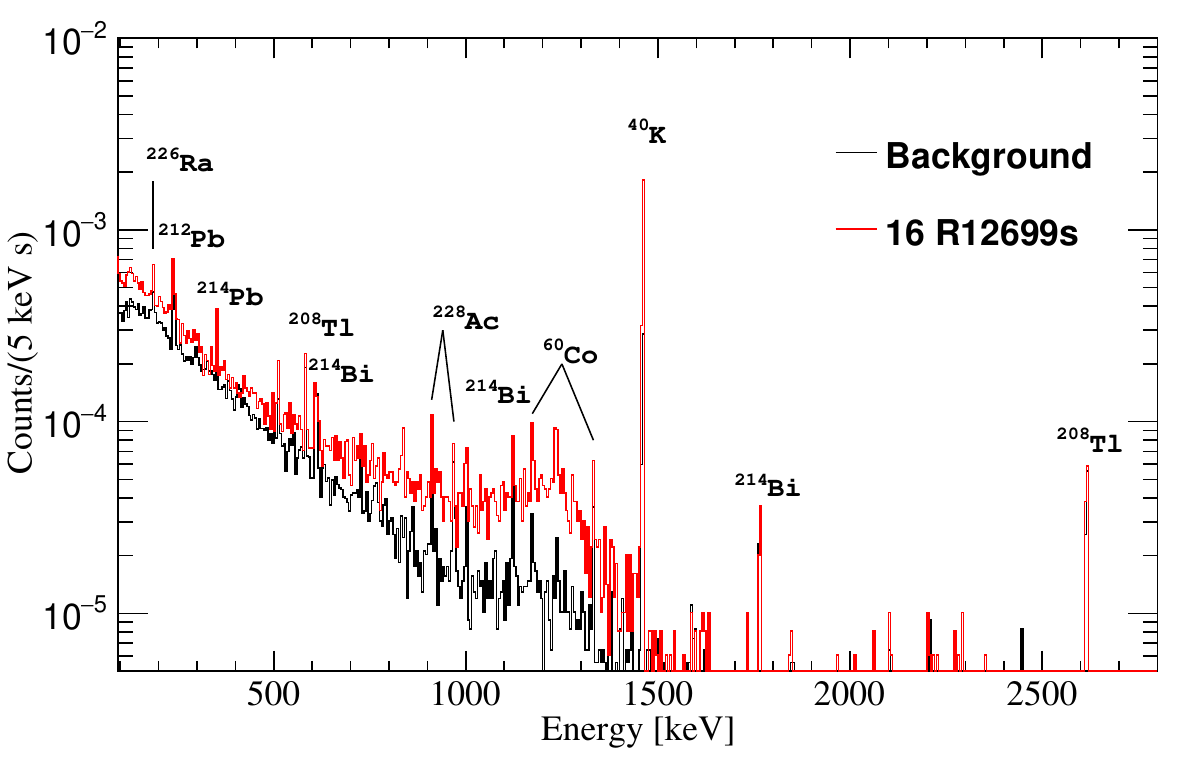}
    \caption{(Left) The measurement arrangement and simulated configuration in Geant4 of PMTs. The crystal, along with its shell, has a cylindrical shape surrounded by the PMTs which are red in color and the Teflon holder is white. (Right) The background and signal spectra for 16 pieces of R12699 PMTs.}
    \label{fig_measure}
\end{figure}

The HPGe detector is located in the China JinPing Underground Laboratory (CJPL), which is the deepest underground laboratory in the world~\cite{CJPL}. Benefiting from the 2400-m rock overburden, the cosmic ray flux at CJPL can reach (3.53 $\pm$ 0.22$_{stat.}$ $\pm$ 0.07$_{sys.}$) $\times$ 10$^{-10}$~cm$^{-2}$s$^{-1}$~\cite{CJPLMuon}, resulting in extremely low background for the HPGe detector. To further enhance the Minimum Detectable Activity (MDA)~\cite{MDA} of the HPGe detector, which is influenced by both background level and detection efficiency, a combination of 10-cm-thick low-background oxygen-free copper and 20-cm-thick lead is employed as shielding~\cite{paper_wxm,thesis_yyk}. 
The geometry of the samples and their constituent materials may be complex, making it challenging to calculate detection efficiencies directly. To obtain precise detection efficiencies of HPGe detector, BambooMC~\cite{BambooMC}, a Monte Carlo simulation framework built on Geant4, is employed. To improve detection efficiency, 16 PMTs are arranged around the crystal region of the HPGe detector, with 8 positioned on the sides and 8 on the top. The PMTs are mounted on Teflon holders, of which background contribution is negligible. The measurement arrangement and its simulated configuration are depicted in Fig.~\ref{fig_measure}. Ultimately, the activities of the radioactive isotope are measured by calculating a weighted average of the gamma peaks with high branching ratios (marked in the energy spectrum shown in Fig.~\ref{fig_measure}). After two weeks of measurement, the MDA can reach about 0.07 mBq/PMT for $^{238}$U(l) and 0.08 mBq/PMT for $^{232}$Th(l).

All materials used in R12699 PMT are screened by the HPGe detector. To better understand the radioactive sources and composition of the R12699 PMT, Tab.~\ref{tab:material} summarizes the measurement results in units of mBq/PMT. 
Aluminum ring (part 2) and ceramic insulator (part 10) are not included in Tab.~\ref{tab:material} as their minimal usage results in negligible levels of radioactivity. And the alloy (part 9) which is used for the electrodes has multiple batches, with all their screening results displayed.

\begin{table}[htbp]
\centering
\caption{The activity of materials used in R12699 PMT (unit: mBq/PMT). $\rm^{232}Th(e)$, $\rm^{232}Th(l)$, $\rm^{238}U(e)$ and $\rm^{238}U(l)$ represents the $\rm^{232}Th$ and $\rm^{238}U$ early and late chain which was described in Sec.~\ref{RI_bulk}. The number in the bracket represents the errors of the activity: 0.67(14) means 0.67$\pm$ 0.14 that is the same in Tab.~\ref{tab:material_replace}, Tab.~\ref{tab3} and Tab.~\ref{tab:pmt_comparsion}. }
\label{tab:material}
\smallskip
\begin{adjustbox}{width=\textwidth, totalheight=\textheight, keepaspectratio}
\begin{tabular}{cccccccccc}
\hline
Part &Material &$\rm ^{60}Co$ &$\rm{^{137}Cs}$ & $\rm^{40}K$ &$\rm^{232}Th(e)$ &\rm $\rm^{232}Th(l)$ &$\rm^{235}U$ &$\rm^{238}U(e)$ &$\rm^{238}U(l)$ \\
\hline
1 &Quartz &\textless0.05 &\textless0.05 &\textless0.71 &\textless0.21 &\textless0.04 &\textless0.76 &\textless1.87 &\textless0.13 \\
\hline
3 &Kovar &0.03(1) &\textless0.01 &\textless0.05 &\textless0.02 &\textless0.01 &\textless0.16 &0.09 &\textless0.01 \\
\hline
4 &Kovar &0.11(1) &\textless0.03 &\textless0.19 &\textless0.06 &\textless0.03 &\textless0.61 &\textless0.36 &\textless0.05 \\
\hline
5 &Kovar &0.22(3) &\textless0.08 &\textless0.46 &\textless0.20 &\textless0.06 &\textless1.51 &\textless0.57 &\textless0.10 \\
\hline
6 &Glass-1 &\textless0.02 &\textless0.01 &0.67(14) &0.13(3) &0.11(1) &\textless0.35 &1.69(17) &0.64(2)\\
\hline
7 &SS &\textless0.01 &- &\textless0.03 &\textless0.01 &\textless0.01 &\textless0.13 &\textless0.04 &\textless0.01 \\
\hline
8 &SS &\textless0.01 &- &\textless0.01 &\textless0.01 &\textless0.01 &\textless0.06 &\textless0.02 &\textless0.01 \\
\hline
9 &Alloy &\textless0.07 &\textless0.08 &\textless0.76 &\textless0.23 &\textless0.06 &\textless0.79 &\textless0.74 &\textless0.15\\
  &        &\textless0.06 &\textless0.06 &\textless0.55 &\textless0.20 &\textless0.09 &\textless0.43 &\textless2.65 &\textless0.11\\
  &        &\textless0.06 &\textless0.05 &\textless0.56 &\textless0.15 &\textless0.05 &\textless1.47 &\textless1.70 &\textless0.13\\
  &        &\textless0.07 &\textless0.08 &\textless0.61 &\textless0.30 &\textless0.06 &\textless1.15 &\textless0.84 &\textless0.16\\
  &        &\textless0.05 &\textless0.07 &\textless0.55 &\textless0.15 &\textless0.05 &\textless0.61 &\textless1.50 &\textless0.14\\
  &        &\textless0.06 &\textless0.07 &\textless0.84 &\textless0.26 &\textless0.06 &\textless1.68 &\textless0.65 &\textless0.07\\
\hline
11 &Glass-2 &- &- &\textless0.01 &\textless0.01 &\textless0.01 &- &\textless0.02 &\textless0.01 \\
\hline
\end{tabular}
\end{adjustbox}
\end{table}

\begin{figure}[!htbp]
\centering
\includegraphics[width=0.7\textwidth]{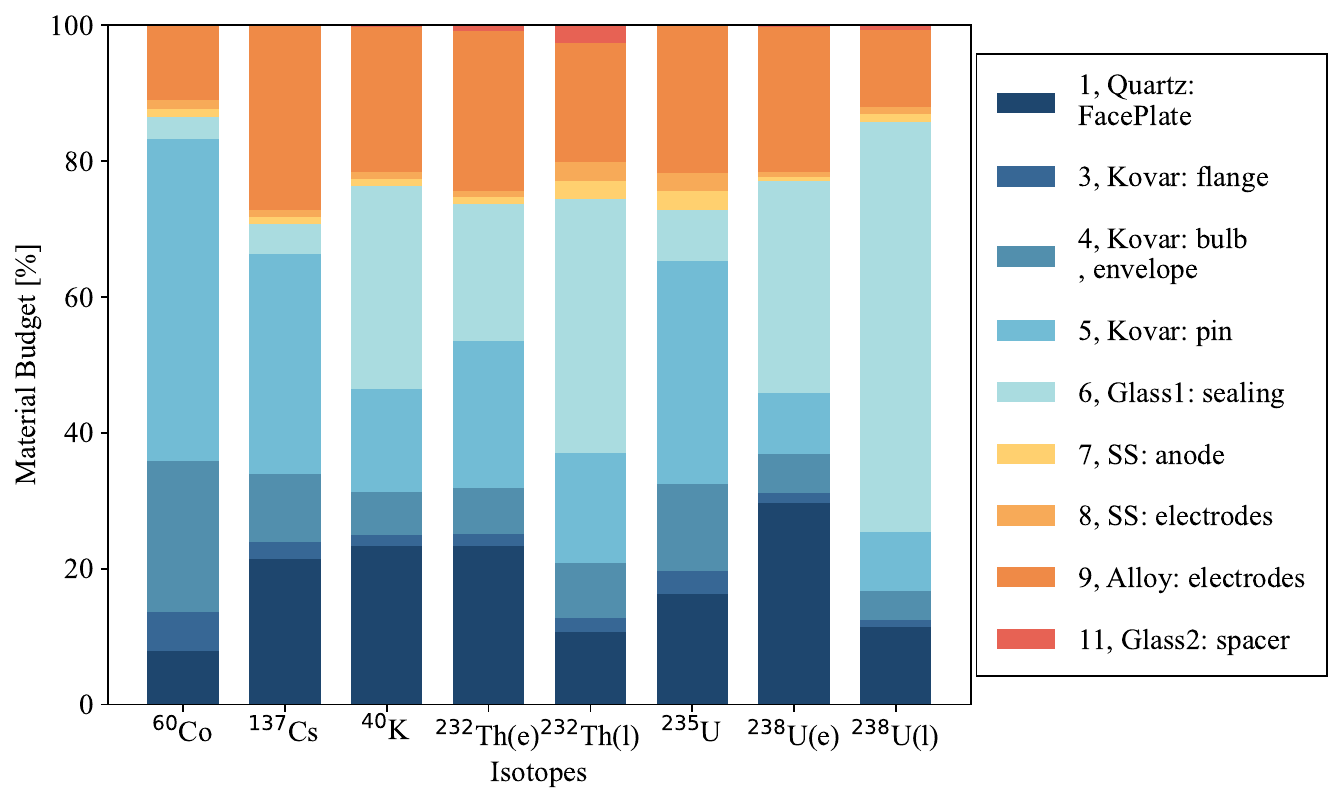}
\caption{The material radioactivity budget of R12699 PMT. }\label{fig_budget}
\end{figure}

Fig.~\ref{fig_budget} visually displays the percentage contribution of each component. For part 9, an average value from different batches of alloy used in various electrode components is applied. From Fig.~\ref{fig_budget}, it is evident that the Kovar metal plate, pin, and pipe (parts 3, 4, 5) are the principal contributors to $\rm^{60}Co$ activity, while the highly radioactive Glass-1 used for sealing significantly contributes to  $\rm^{238}U(l)$ and $\rm^{232}Th(l)$ activities.

To reduce radioactivity while preserving PMT electrical performance, it is necessary to replace both the Kovar and glass materials. After a thorough evaluation of various alternatives, the replacement materials have been identified and their radioactivities are measured (Tab.~\ref{tab:material_replace}). 
By replacing Kovar which is an iron-cobalt-nickel alloy with a new iron-nickel alloy, the $^{60}\rm Co$ radioactivity has been reduced.
Meanwhile, the radioactivity of the new low-background glass leads to a roughly 9 times reduction in $\rm^{238}U(l)$ radioactivity and a 4 times decrease in $\rm^{232}Th(l)$ radioactivity. The raw materials for the new glass are carefully selected to be low-radioactive, and a platinum crucible is used to prevent contamination of radioactive isotopes, such as $\rm^{232}Th(l)$, $\rm^{238}U(l)$, and others.

\begin{table}[htbp]
\centering
\caption{The activities of PMT materials for replacement (unit:mBq/PMT). }
\label{tab:material_replace}
\smallskip
\begin{adjustbox}{width=\textwidth, totalheight=\textheight, keepaspectratio}
\begin{tabular}{cccccccccc}
\hline
Part &\makecell{Material} &$\rm ^{60}Co$ &$\rm{^{137}Cs}$ & $\rm^{40}K$ &$\rm^{232}Th(e)$ &\rm $\rm^{232}Th(l)$ &$\rm^{235}U$ &$\rm^{238}U(e)$ &$\rm^{238}U(l)$ \\
\hline
3 &Alloy &\textless0.004 &\textless0.005 &\textless0.055  &\textless0.017 &\textless0.004 &\textless0.11 &\textless0.043 &\textless0.005 \\
\hline
4 &Alloy &\textless0.014 &\textless0.02 &\textless0.21 &\textless0.07 &\textless0.014 &\textless0.42 &\textless0.16 &\textless0.018 \\ 
\hline
5 &Alloy &\textless0.04 &\textless0.06 &\textless0.67 &\textless0.21 &\textless0.05 &\textless1.34 &\textless0.52 &\textless0.06\\
\hline
6 &New glass &- &\textless0.01 &0.56(11) &\textless0.01 &0.03(1) &0.02(1) &0.18(5) &0.07(1)\\
\hline
\end{tabular}
\end{adjustbox}
\end{table}

Three versions of the PMTs were produced during the iterative development processes: v0 (original), v1 (with Kovar replaced), and v2 (with both Kovar and Glass-1 replaced). To verify the effect of these modifications, measurements were conducted across multiple versions of PMTs, with the results presented in Tab.~\ref{tab3}. Multiple batches of the three PMT versions were screened, and average results were calculated for the comparison between different versions. From v0 to v1, the radioactivity of $^{60}$Co is reduced from 0.7~mBq/PMT to about 0.08~mBq/PMT. Moreover, the activity of $\rm^{238}U(l)$ in v2 has been lowered to 0.06~mBq/PMT and the activity of $\rm^{232}Th(l)$ has been reduced to 0.09~mBq/PMT. 

\begin{table}[H]
\centering
\caption{Screening results of three versions of PMTs(unit: mBq/PMT).}
\label{tab3}
\begin{adjustbox}{width=\textwidth, totalheight=\textheight, keepaspectratio}
\begin{tabular}{ccccccccccc}
\hline
Version & Quantity &$\rm ^{60}Co$ &$\rm{^{137}Cs}$ & $\rm^{40}K$ &$\rm^{232}Th(e)$ &\rm $\rm^{232}Th(l)$ &$\rm^{235}U$ &$\rm^{238}U(e)$ &$\rm^{238}U(l)$\\
\hline
v0 &4 &0.76(10) &\textless0.16 &26(2) &\textless0.46 &\textless0.12 &\textless1.11 &\textless6.35 &0.60(14) \\
v0&3 &0.74(9) &\textless0.07 &31(2) &\textless0.26 &\textless0.20 &\textless0.15 &\textless3.88 &\textless0.50 \\
v0 &4 &1.01(10) &\textless0.20 &32(2) &\textless0.26 &\textless0.64 &\textless0.68 &\textless5.05 &0.61(15) \\
v0 &4 &\textless0.26 &\textless0.22 &37(3) &\textless0.63 &\textless1.30 &\textless0.40 &\textless3.92 &\textless0.69 \\
v0 &4 &0.69(12) &\textless0.08 &26(2) &\textless0.32 &\textless0.16 &\textless0.18 &\textless5.78 &\textless0.52 \\
average &- &0.66(5) &\textless0.08 &30(1) &\textless0.20 &\textless0.37 &\textless0.32 &\textless3.09 &0.41(8) \\
\hline
v1 &2 &\textless0.15 &\textless0.25 &37(2) &\textless0.92 &\textless1.28 &\textless0.28 &\textless9.44 &\textless1.07 \\
v1 &2 &\textless0.26 &\textless0.22 &37(3) &\textless0.63 &\textless1.30 &\textless0.40 &\textless3.92 &\textless0.69 \\
v1 &4 &\textless0.07 &\textless0.05 &38(3) &\textless0.19 &\textless0.14 &\textless0.30 &\textless1.88 &0.42(11) \\
v1 &8 &\textless0.07 &\textless0.09 &31(2) &\textless0.40 &\textless0.40 &\textless0.48 &\textless1.03 &0.47(11) \\
v1 &16 &0.08(2) &\textless0.06 &41(5) &\textless0.19 &\textless0.23 &\textless0.23 &\textless3.73 &0.43(4) \\
v1 &16 &0.09(2) &\textless0.13 &37(5) &0.45(10) &0.34(8) &\textless0.17 &\textless1.97 &0.39(4) \\
v1 &16 &0.12(3) &\textless0.65 &38(5) &0.33(10) &0.29(8) &\textless0.42 &\textless0.78 &0.41(4) \\
v1 &16 &\textless0.04 &\textless0.06 &34(2) &\textless0.28 &0.23(7) &\textless0.28 &\textless2.38 &0.39(6) \\
average &- &\textless0.08 &\textless0.12 &37(1) &\textless0.31 &0.34(7) &\textless0.14 &\textless1.91 &0.43(5) \\
\hline
v2 &16 &0.13(2) &\textless0.07 &41(5) &\textless0.07 &\textless0.10 &\textless0.05 &\textless1.16 &0.10(3) \\
v2 &16 &0.07(2) &\textless0.05 &31(2) &0.24(7) &\textless0.14 &\textless0.21 &\textless1.87 &\textless0.11 \\
v2 &16 &0.07(2) &\textless0.05 &32(2) &0.22(7) &0.17(5) &\textless0.18 &\textless1.27 &\textless0.08 \\
v2 &16 &0.06(2) &\textless0.06 &36(2) &\textless0.21 &\textless0.21 &\textless0.25 &\textless1.61 &\textless0.17 \\
average &- &0.08(1) &\textless0.04 &35(2) &0.14(3) &0.09(3) &\textless0.11 &\textless1.05 &0.06(2) \\
\hline
\end{tabular}
\end{adjustbox}
\end{table}

\begin{figure}[!ht]
\centering
\includegraphics[width=0.7\textwidth]{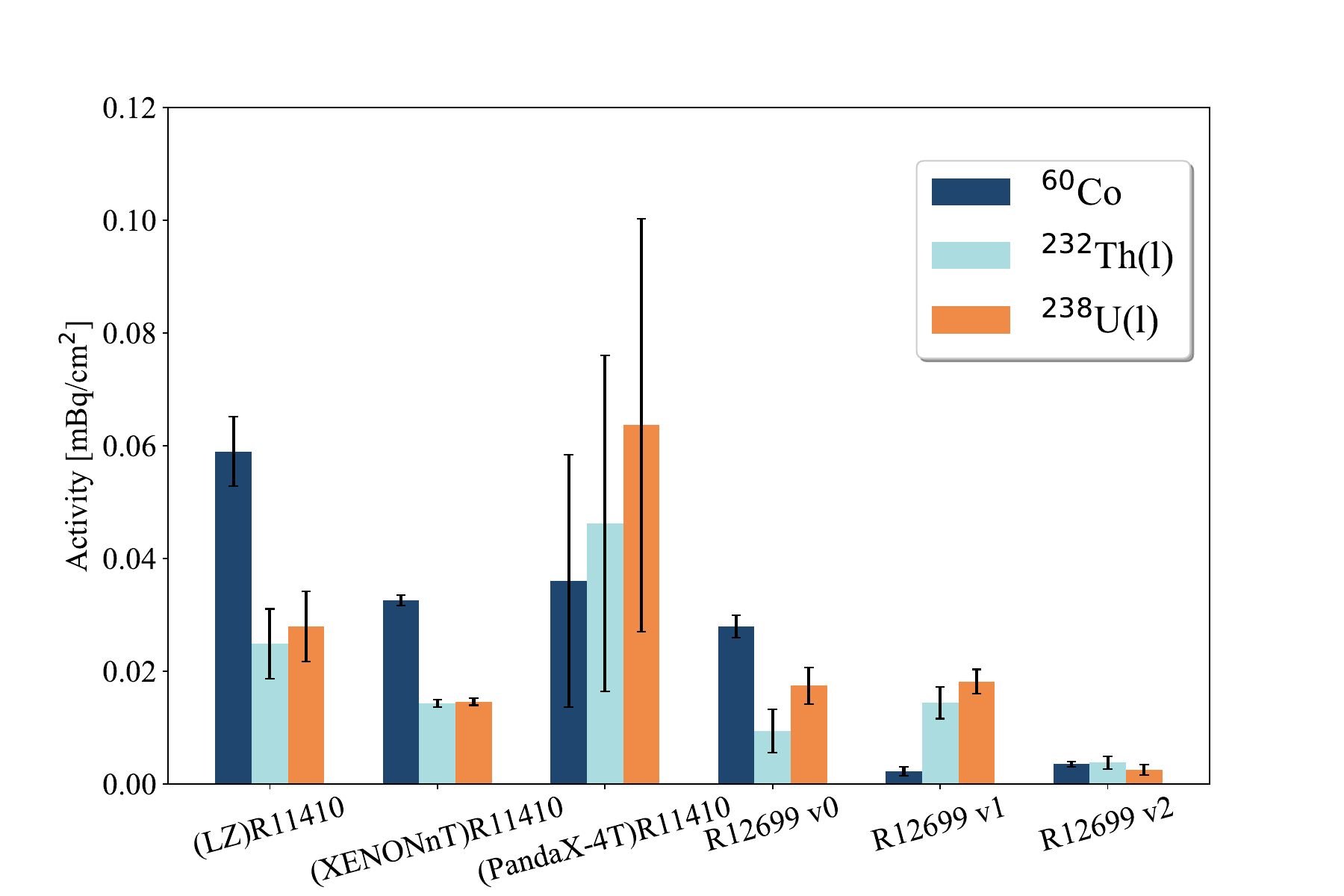}
\caption{The radioactivity comparison of PMTs (Unit: mBq/$\rm cm^{2}$).}\label{PMT_compare}
\end{figure}

Tab.~\ref{tab:pmt_comparsion} provides a summary of radioactivity of PMTs employed in rare event search experiments. The radioactivity of R12699 PMT shows a decrease of more than 15 times for $\rm^{60}Co$, $^{232}$Th(l) and $^{238}$U(l) compared to that of PMT R11410 used in PandaX-4T. It can be indicated from Fig.\ref{PMT_compare} that the radioactivity of $\rm^{60}Co$ is one order of magnitude better than that of LZ and XENONnT PMTs. Additionally, the radioactivity of $\rm^{238}U(l)$ is 5 to 9 times better than them. The $\rm^{60}Co$ and $\rm^{238}U(l)$ radioactivities of R12699 PMT are similar to that of silicon photomultipliers~\cite{S13371,SiPM,NEXT}. 

\begin{table}[!ht]
\centering
\caption{Screening results for PMTs used in various experiments. }
\label{tab:pmt_comparsion}
\begin{adjustbox}{width=\textwidth, totalheight=\textheight, keepaspectratio}
\begin{tabular}{ccccccccc}
\hline
\makecell{PMT\\(mBq/cm$^2$)} &$\rm ^{60}Co$ &$\rm{^{137}Cs}$ & $\rm^{40}K$ &$\rm^{232}Th(e)$ &\rm $\rm^{232}Th(l)$ &$\rm^{235}U$ &$\rm^{238}U(e)$ &$\rm^{238}U(l)$\\
\hline
\makecell{R11410-10\\(LZ)\cite{LZbkg}} & 0.059(6) & - & 0.38(3) & 0.043(25) & 0.025(6) & 0.025(19) & 0.15(7) & 0.028(6) \\
\hline
\makecell{R11410-20\\(XENONnT)\cite{XENONbkg}} & 0.033(1) & \textless0.004 & 0.44(2) & 0.015(2) & 0.014(1) & 0.012(3) & 0.28(6) & 0.015(2) \\
\hline
\makecell{R11410-23\\(PandaX-4T)\cite{PandaX-4Tbkg}} & \textless0.073 & \textless0.057 & \textless0.69 & \textless0.24 & \textless0.095 & \textless0.88 & \textless1.75 & \textless0.12 \\
\hline
\makecell{R12699-406-M4(v2)\\(this work)} &0.004(1) &\textless0.002 &1.5(1) &0.006(1) &0.004(1) &\textless0.005 &\textless0.045 &0.003(1) \\
\hline
\makecell{R13111\\(XMASS)\cite{XMASS}} & 0.003(1) & - & 0.052(13) & 0.005(2) & - & - & \textless0.036 & 0.011(2) \\
\hline
\end{tabular}
\end{adjustbox}
\end{table}

\subsection{Radon emanation rates}
Radon and its decay daughters are significant sources of background in rare event search experiments. $^{222}$Rn originates from the decay of $^{226}$Ra present in the detector materials and will emanate into xenon. The PMTs are close to the xenon active volume thus their radon emanation rate needs to be carefully screened and controlled.
To better control the background of the experiment, the $^{222}\rm Rn$ emanation rates of the three versions of R12699 PMTs are measured by detecting the radon daughters $\rm ^{214}Po$ ($\rm T_{1/2}$ = 164~$\mu$s) and $\rm ^{218}Po$ ($\rm T_{1/2}$ = 3.1~min) due to their short half-lifes. 
Since most of the $\rm ^{214}Po$ and $\rm ^{218}Po$ are positively charged, they can be collected using an electric field to the detecting surface and their decay alpha particles can be detected.
In PandaX-4T, two radon emanation systems based on this electrostatic collection method are established to measure it~\cite{PandaX-4Tbkg}. The screening results are shown in Tab.\ref{tab4}. 

\begin{table}[htbp]
\centering
\caption{Radon emanation results (unit: $\rm \mu$Bq/PMT).}
\label{tab4}
\begin{adjustbox}{width=\textwidth, totalheight=\textheight, keepaspectratio}
\begin{tabular}{cccccccccc}
\hline
Type &\makecell{R11410\\(LZ)}&\makecell{R11410\\(XENONnT)} &\makecell{R12699\\(v0)} &\makecell{R12699\\(v1)}  & \makecell{R12699\\(v1)}& \makecell{R12699\\(v1)}& \makecell{R12699\\(v2)}&\makecell{R12699\\(v2)}&\makecell{R12699\\(v2)} \\
\hline
 Batch& -& -& -&  1& 2& 3& 1&2&3\\
\hline
Rate &
$\rm 1900^{+1700}_{-1900}$ &
\makecell{2±1} &
\makecell{\textless11.5} &
\makecell{\textless1.3} &
\makecell{\textless3.5} &
\makecell{\textless2.68} &
\makecell{\textless3.0} &
\makecell{\textless3.3} &
\makecell{\textless2.1} \\
\hline
\end{tabular}
\end{adjustbox}
\end{table}

\subsection{Surface radioactivity}
The surface alpha radioactivity can induce neutron backgrounds in the detector via the ($\rm \alpha$,n) reaction. $^{210}$Po is measured because the long half-life of $^{210}$Pb allows it to deposit on the sample surface over time, and in the decay chain, only $^{210}$Po emits alpha and has a short half-life among the subsequent isotopes. The surface radioactivity of the PMT quartz window is measured by a commercial Alpha Mega system from ORTEC. In order to remove surface activity, an efficient surface cleaning method is necessary. Wiping with alcohol is used in the cleaning procedure.
To verify the cleaning effect, the quartz sample used in PMT window was first exposed to a high-radon environment for about half a year.
It can be obtained from Tab.~\ref{tab:my_label} that after two rounds of cleaning procedures, the surface activity of the quartz sample is at the same level as that of the initial case, which means the polluted $\rm ^{210}Po$ is removed.
\begin{table}[H]
    \centering
    \caption{Cleaning effect of quartz (unit: $\rm \mu Bq/cm^{2}$).}
    \begin{tabular}{c|c|c|c|c} \hline
        Samples &The initial & The polluted & The first clean & The second clean\\ 
        \hline
        1 & \textless 18.4 & 236.9 $\pm$ 27.8 & 27.8 $\pm$ 8.1 & \textless 22.5\\ 
        \hline
    \end{tabular}
    \label{tab:my_label}
\end{table}

    \section{Cryogenic test setups}
    \label{sec:setups}

    The experimental setups include the PMT readout base and the cryogenic testing platform, which will be introduced in this section.

    \subsection{The PMT readout base design}
    \label{subsec:HV_base}

\begin{figure}[htbp]
		\centering
		\includegraphics[width=\textwidth]{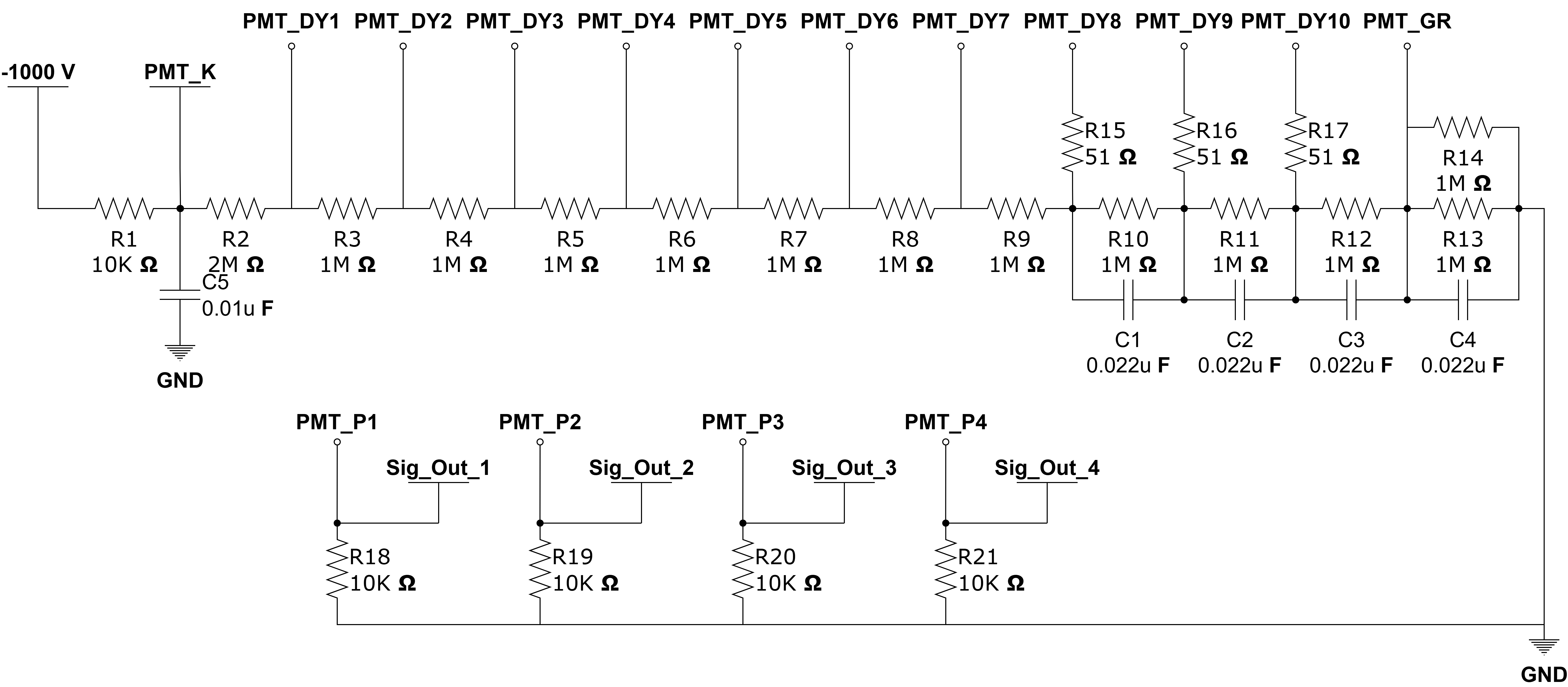}
		\caption{The schematic diagram of the -1000 V biased PMT readout base with the producer recommended distribution ratio. K represents for the cathode, DY refers to the dynode, P represents for the anode, and GR denotes the guard ring.        }
        \label{fig:schematic}
    \end{figure}

    The readout base distributes bias voltages to the PMT's cathode, dynodes, and anodes, as shown in Fig.~\ref{fig:schematic}, and extract PMT signals from the anodes, which are then outputted to subsequent electronic systems. These bases are designed to accommodate nine tubes on a $210~\rm{mm} \times 210~\rm{mm}$ square board, forming a $3\times 3$ array, while each high voltage (HV) of a PMT can be adjusted and each signal of the four channels on a PMT can be read out individually. This design of integrated base serves for PMT batch testing.

    We have designed both positive HV biased readout bases and negative HV biased ones, and both types work properly during thermal cycles from room temperature to -100~$^{\circ}$C. Due to the limited area available for the PMT readout base layout, we use the negative HV-biased readout base for the batch test. This design, with fewer components and larger spacing between them, reduces the risk of HV trips.

    \begin{table}[htbp]
        \caption{The original and optimized voltage distribution ratio.}
        \label{tab:base_ratio}
        \centering
        \resizebox{\textwidth}{!}{
        \begin{tabular}{p{0.116\textwidth}@{}p{0.034\textwidth}@{}p{0.034\textwidth}@{}p{0.034\textwidth}@{}p{0.034\textwidth}@{}p{0.034\textwidth}@{}p{0.034\textwidth}@{}p{0.034\textwidth}@{}p{0.034\textwidth}@{}p{0.034\textwidth}@{}p{0.034\textwidth}@{}p{0.034\textwidth}@{}p{0.034\textwidth}@{}p{0.034\textwidth}@{}p{0.034\textwidth}@{}p{0.034\textwidth}@{}p{0.034\textwidth}@{}p{0.034\textwidth}@{}p{0.034\textwidth}@{}p{0.034\textwidth}@{}p{0.034\textwidth}@{}p{0.034\textwidth}@{}p{0.034\textwidth}@{}p{0.034\textwidth}@{}p{0.034\textwidth}@{}p{0.034\textwidth}@{}p{0.034\textwidth}}
        \hline
        \multicolumn{1}{|c|}{Electrodes} & \multicolumn{2}{c|}{K} &\multicolumn{2}{c|}{Dy1} &\multicolumn{2}{c|}{Dy2}& \multicolumn{2}{c|}{Dy3}& \multicolumn{2}{c|}{Dy4}& \multicolumn{2}{c|}{Dy5}& \multicolumn{2}{c|}{Dy6}& \multicolumn{2}{c|}{Dy7}& \multicolumn{2}{c|}{Dy8}& \multicolumn{2}{c|}{Dy9}& \multicolumn{2}{c|}{Dy10}& \multicolumn{2}{c|}{GR} & \multicolumn{2}{c|}{P}\\
        \hline
        \multicolumn{2}{|c|}{Original}&\multicolumn{2}{c|}{2}& \multicolumn{2}{c|}{1}&\multicolumn{2}{c|}{1}&\multicolumn{2}{c|}{1}&\multicolumn{2}{c|}{1}&\multicolumn{2}{c|}{1}&\multicolumn{2}{c|}{1}&\multicolumn{2}{c|}{1}&\multicolumn{2}{c|}{1}&\multicolumn{2}{c|}{1}&\multicolumn{2}{c|}{1}&\multicolumn{2}{c|}{0.5}&\\
        \multicolumn{2}{|c|}{Optimized}&\multicolumn{2}{c|}{1.5}&\multicolumn{2}{c|}{1.1}&\multicolumn{2}{c|}{1.4}&\multicolumn{2}{c|}{1}&\multicolumn{2}{c|}{1}&\multicolumn{2}{c|}{1}&\multicolumn{2}{c|}{1}&\multicolumn{2}{c|}{1}&\multicolumn{2}{c|}{1}&\multicolumn{2}{c|}{1}&\multicolumn{2}{c|}{1}&\multicolumn{2}{c|}{0.5}&\\
        \cline{1-26}
         & & & & & & & & & & & & & & & & & & & & & & & & & &
        \end{tabular}
        }
    \end{table}
    
    The voltage distribution ratio has been optimized from the Hamamastu's recommended value to achieve a higher PMT gain value, as shown in Tab.~\ref{tab:base_ratio}. According to Ref.~\cite{f}, the gain improvement calculated from the optimization is around 38\%, and the tested gain improvement is about 30\%, as shown in Tab.~\ref{tab:G_improve}. However, the PMT electron collection efficiency (CE) may also be influenced as the voltage between the cathode and the first dynode is reduced from 160~V to 120~V (under -1000~V biased HV) by the optimization. We have measured the photoelectron (PE) count of the same PMT under identical luminous conditions for both distribution ratios and found that the electron CE with the optimized ratio has a negligible decrease by less than 5\% compared to the original configuration.

    \begin{table}[htbp]
        \caption{The gain comparison of the two R12699 PMTs under both original and optimized distribution ratio.}
        \label{tab:G_improve}
        \centering
        \resizebox{\textwidth}{!}{
        \begin{tabular}{c|c|c|c}
        \hline
        PMT No.&Avg gain (optimized distribution ratio)&Avg gain (original distribution ratio)& Gain improvement\\
        \hline
        MA0053&$4.69\times 10^6$&$3.59\times 10^6$&30.6\%\\
        \hline
        MA0038&$5.02\times 10^6$&$3.69\times 10^6$&36\%\\
        \hline
        \end{tabular}
        }
    \end{table}

\begin{figure}[htbp]
        \centering
        \includegraphics[width=0.7\textwidth]{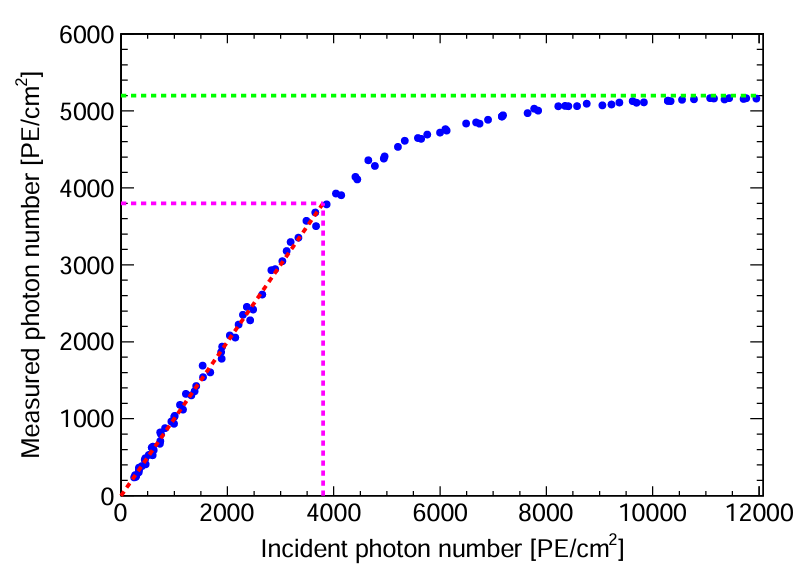}
        \caption{The curve of incident PE number per unit area and the PMT measured PE number per unit area. The red dotted line is the ideal linear curve, and the magenta dotted lines mark the point that the curve start to deviate from linearity, at about 3800~PE/cm$^2$. The green dotted line corresponds to the upper limit of PMT measured PE number.}
        \label{fig:saturation}
    \end{figure}

    The saturation limit of the R12699 PMTs on the readout bases is also measured, by illuminating the PMT with various light intensity and calculating the PE count from the corresponding PMT waveform~\cite{g}. The measured R12699 PMTs reach saturation when the cathodes produce approximately 30,000~PEs per channel, or 3800~PEs/cm$^2$, as shown in Fig.~\ref{fig:saturation}.

    \subsection{Design of the cryogenic testing platform}
    \label{sec:cryo_setup}
    
\begin{figure}[htbp]
		\centering
		\includegraphics[width=0.7\textwidth]{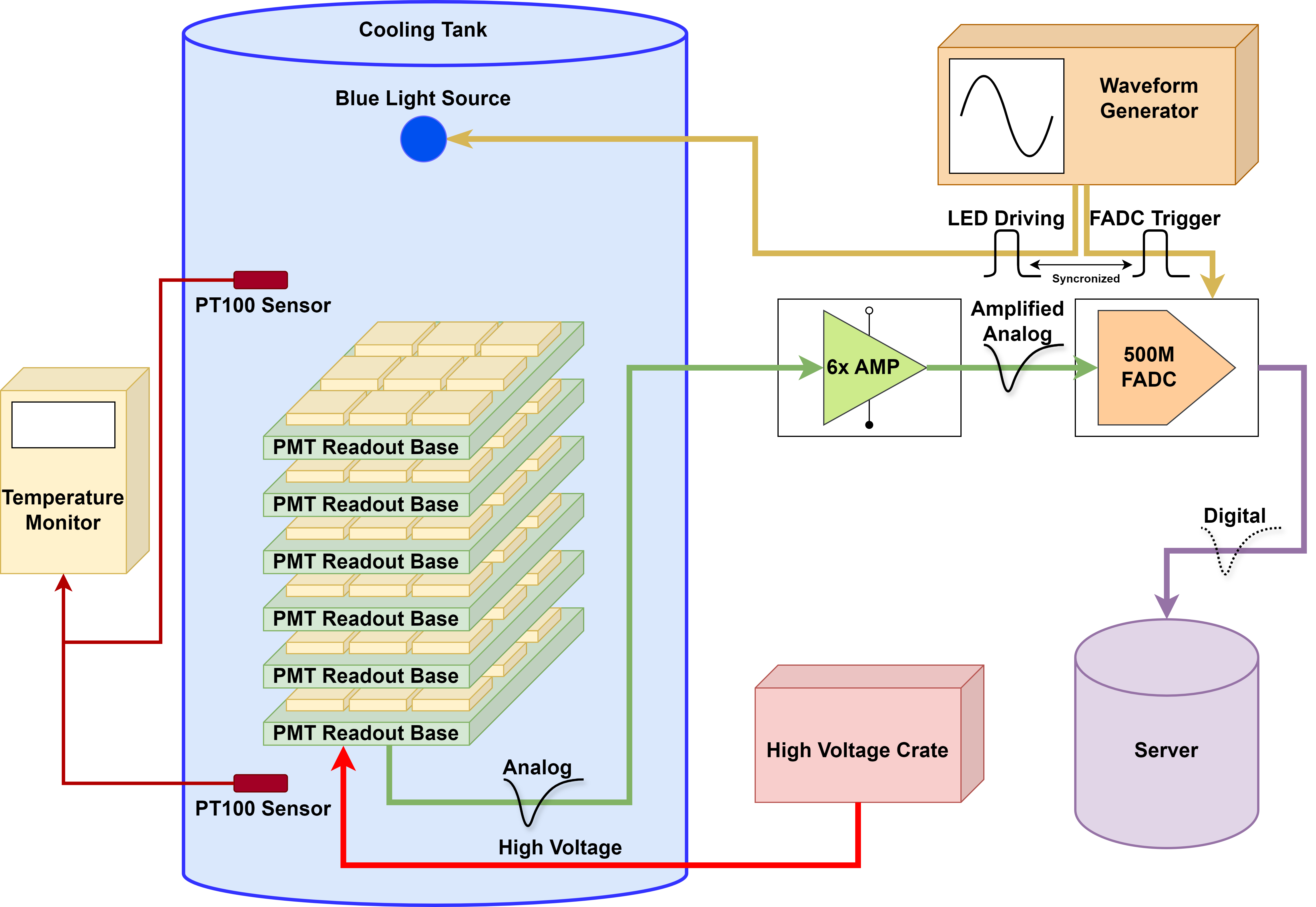}
		\caption{The block diagram illustrating the instruments involved in the PMT-batch qualification, and the route of PMT signals.}
        \label{fig:test_setup} 
    \end{figure}

    A batch of 54 R12699 PMTs has been qualified. Each PMT undergoes three cooling and rewarming cycles to evaluate its cryogenic stability at approximately -100~$^{\circ}$C. The overall setup is illustrated in Fig.~\ref{fig:test_setup}.
    
\begin{figure}[htbp]
		\centering
		\includegraphics[angle=90,width=.8\textwidth]{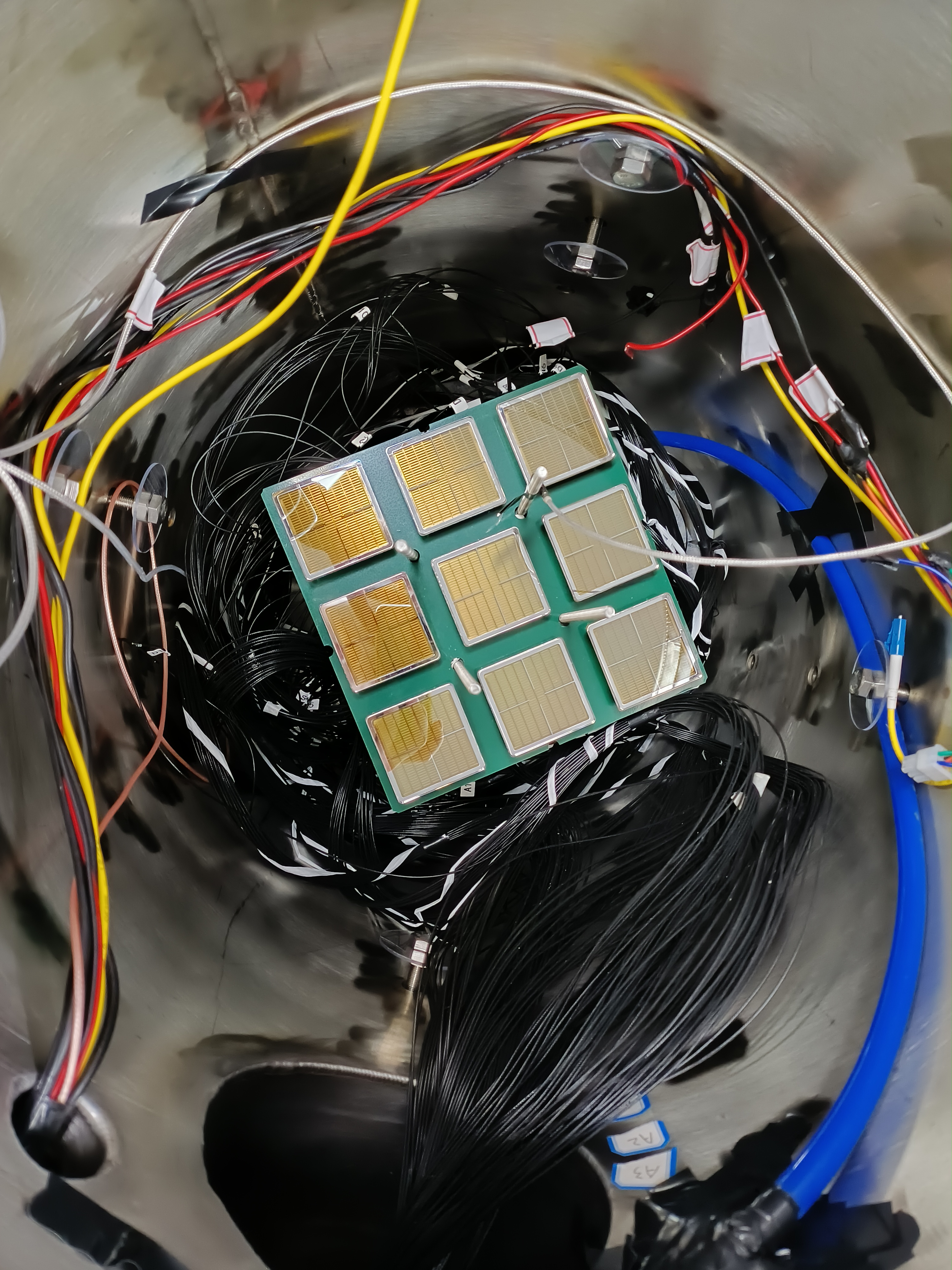}
		\caption{A view of the inner volume of the Dewar vessel, the PMT bases are in the middle and the nitrogen blowpipe (blue, on the top of the figure) are extended to the bottom. }
        \label{fig:cryo_setup} 
    \end{figure}

    The qualified PMTs are placed in a well-grounded, nitrogen-filled Dewar vessel along with their readout bases and a blue light source, cooled by an anhydrous-ethanol-refrigerator to as low as -105~$^{\circ}$C. The PMT readout bases are placed layer by layer, as shown in Fig.~\ref{fig:cryo_setup}, and are connected with outside through coaxial cables and vacuum feedthroughs. Every 9 PMTs are put on the top readout base to be illuminated and measured the gains and AP probabilities.

    The raw signals from the PMTs are amplified by a factor of 6 before being sent to flash analog-to-digital converters (FADC). The customized FADCs, with a 14-bit resolution and a sampling rate of 500 million samples per second (MSPS)~\cite{h}, are self-triggered during DCR measurements. And for gain and AP measurements, the FADCs are triggered by a waveform generator, while the waveform generator also synchronously drives the light source.
    
    During the cryogenic test, the temperature difference between the top readout base and the bottom is about 2~$^{\circ}$C. The nitrogen pressure in the cooling tank is maintained 1~atm higher than the outer air pressure from the beginning of cooling, preventing water vapor from entering and causing short circuit.
    
    \section{PMT electrical performance}
    \label{sec:result}

    This section summarizes the tests conducted on the batch of 54 R12699 PMTs and the results. The measurements were performed using the testing platform described in the previous section, with each PMT following the same procedure to ensure comparable results. The parameters for PMT characterization are described.
    
    \subsection{Gain}
    \label{subsec:gain}

    PMT gain is a crucial characteristic that indicates the PMT's response to incident photons. Higher gain values improve the signal-to-noise ratio (SNR) of the subsequent readout system, which benefits the detection of weak luminescence and enhances sensitivity in the low-energy regions of WIMPs searches.
    
    The PMT gain is defined as the number of electrons collected at the PMT anode resulting from the multiplication induced by a single PE (SPE) emitted from the PMT cathode, and the LED luminosity method is used in our setup to evaluate the gain. Hamamatsu Photonics K.K tests its products by exposing PMTs to a steady light source and measuring the corresponding current with an ammeter. In our qualification, however, to simulate actual operating conditions in liquid xenon detectors as closely as possible, the anode electron numbers are calculated by integrating the PMT signals, with the LED luminosity set to induce SPE events. The FADCs are triggered by pulses synchronized with LED driving signals with a fixed delay to maintain the relative position of the PE waveform within the detection window. Integration within this fixed window helps minimize the noise.
	
    The charge spectrum of a single PMT channel consists of baseline noise, SPE and double-PE signals, and some under-amplified signals. Signals involving 3 PEs and more are negligible under our configuration. The total fit function is expressed as a function of the measured charge $x$:
    \begin{equation*}
		f(x)=n_0\exp{\left(-\frac{(x-\mu_0)^2}{2\sigma_0^2}\right)}+\sum_{i=1}^{2}n_i\exp{\left(-\frac{(x-i\mu_1-\mu_0)^2}{2(i\sigma_1^2+\sigma_0^2)}\right)}+n^{\prime}\exp(-x\alpha)
    \label{equ::3gaus}
    \end{equation*}
    The first Gaussian term, with amplitude $n_0$, mean $\mu_0$, and width $\sigma_0$, represents the baseline noise, forming the pedestal. The second term is a sum of Gaussians representing the SPE and double-PE peaks. The SPE is characterized by Gaussian parameters $n_1$, $\mu_1$, and $\sigma_1$. Double-PE contributions follow Poissonian statistics, with their Gaussian parameters dependent on the SPE values. The final exponential term accounts for under-amplified PE signals, where $n^{\prime}$ denotes the probability of under-amplified PEs and $\alpha$ is the coefficient governing the exponential decrease \cite{s,t}, improving the fit.
    
    \begin{figure}[htbp]
        \centering
        \includegraphics[width=0.45\textwidth]{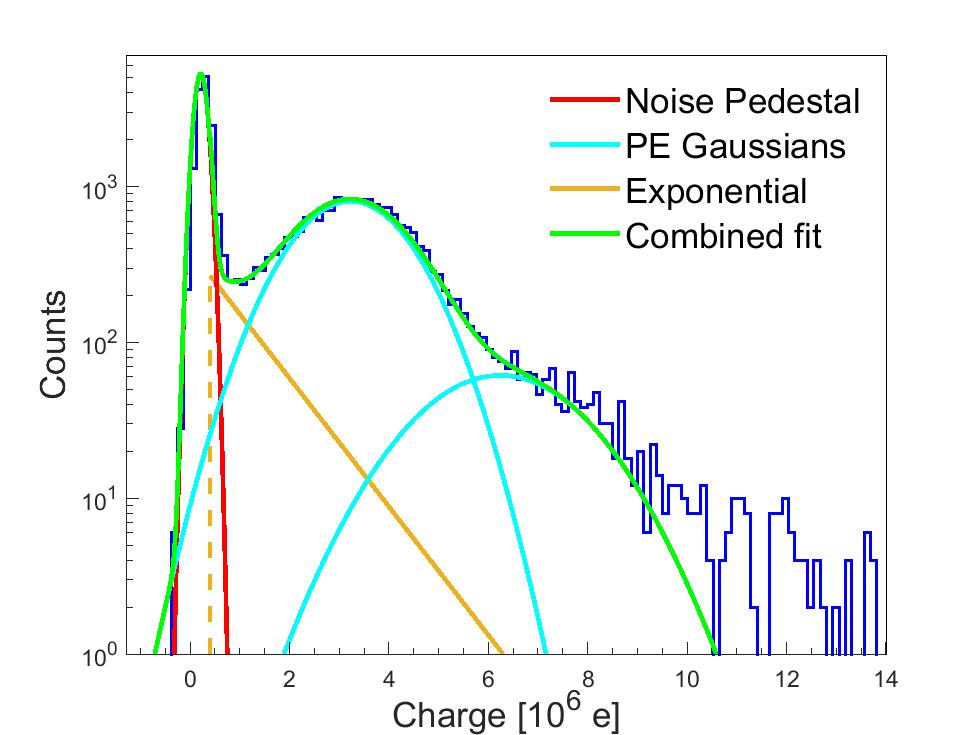}
        \qquad
        \includegraphics[width=0.45\textwidth]{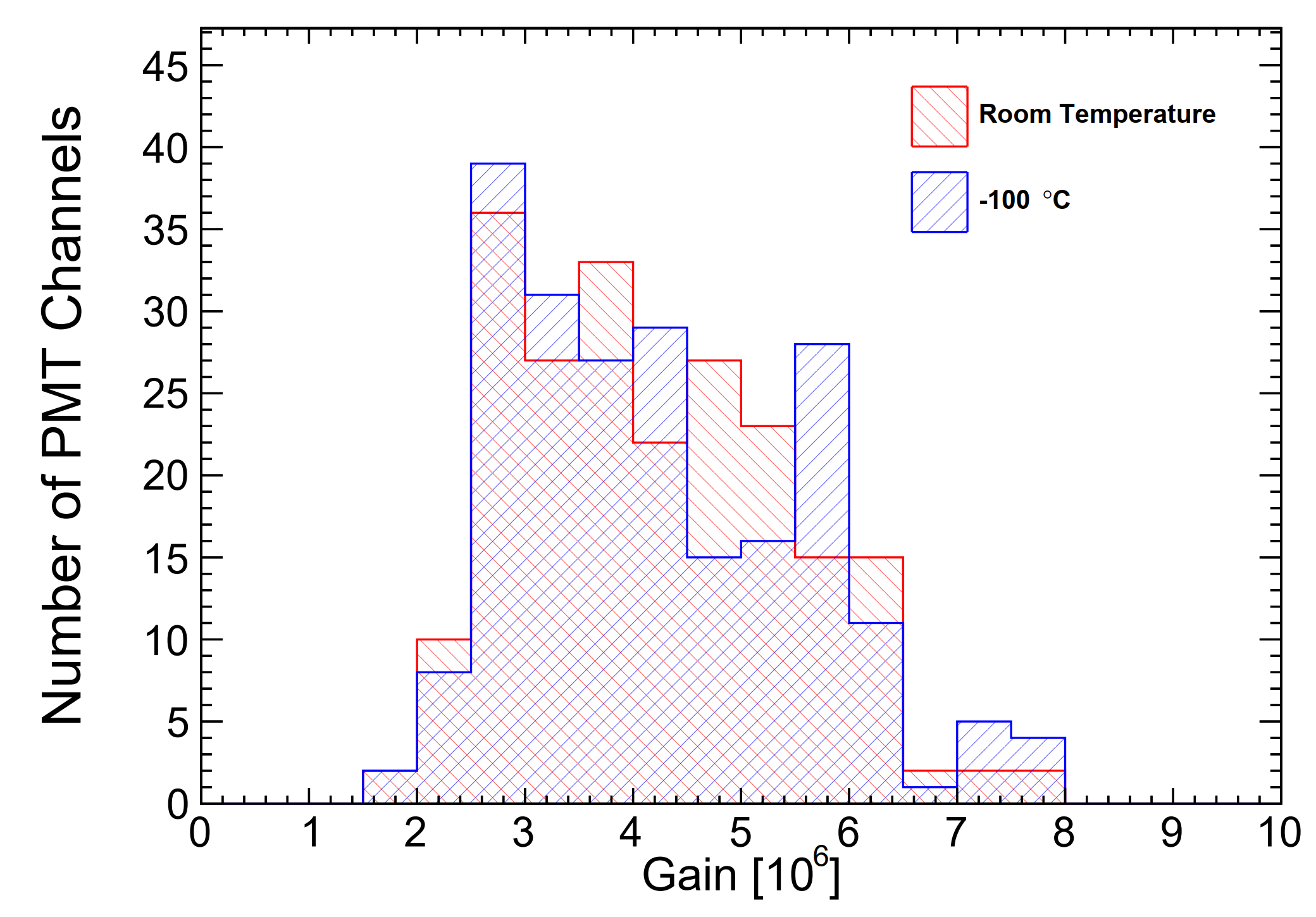}
        \caption{(Left) The charge spectrum of a single PMT channel. The data is shown in blue and the combined fit is shown in green. (Right) Distribution of the measured gains at 1000 V for 54 PMTs (216 channels) in both room temperature (red) and -100~$^{\circ}$C (blue).}
        \label{fig:gain_fit_and_distribute} 
    \end{figure}
    
    The gain is computed as $g = \mu_1 / e$, where $e$ is the elementary charge. An example of the charge spectrum is shown in Fig.~\ref{fig:gain_fit_and_distribute} (Left). Fig.~\ref{fig:gain_fit_and_distribute} (Right) shows the distribution of gains derived from the spectra. The average gain for all PMTs at -1000~V is $4.18 \times 10^6$ at room temperature, and $4.23 \times 10^6$ at -100$^{\circ}$C. There is no significant difference in PMT gain between the two tested temperatures.

    The SPE resolution can also be derived from the spectra, and it is defined as $R = \sigma_1 / \mu_1$, where $\sigma_1$ and $\mu_1$ are the standard deviation and mean of the Gaussian function describing the SPE peak. SPE resolution indicates the ability to distinguish between the noise pedestal and the single-photon response. The average SPE resolution across all tested PMT channels is 42\%.
	
    \subsection{Dark count rates}
    \label{subsec:DCRs}
	
    DCR of a PMT characterizes the occurring frequency of signals which have an amplitude higher than a given threshold in the absence of the light source. Thermal electron emission makes the most contribution to the dark counts, which is significantly mitigated at cryogenic temperatures. Alternative sources, including electron field emission (stemming from high bias voltages) as well as radioactivity and cosmic particle interactions also produce dark signals. A small DCR is crucial for minimizing coincidental events in a liquid xenon experiment, as such events can lead to erroneous associations with the detection of scintillation signals within the detector.
	
\begin{figure}[htbp]
        \centering
        \includegraphics[width=0.45\textwidth]{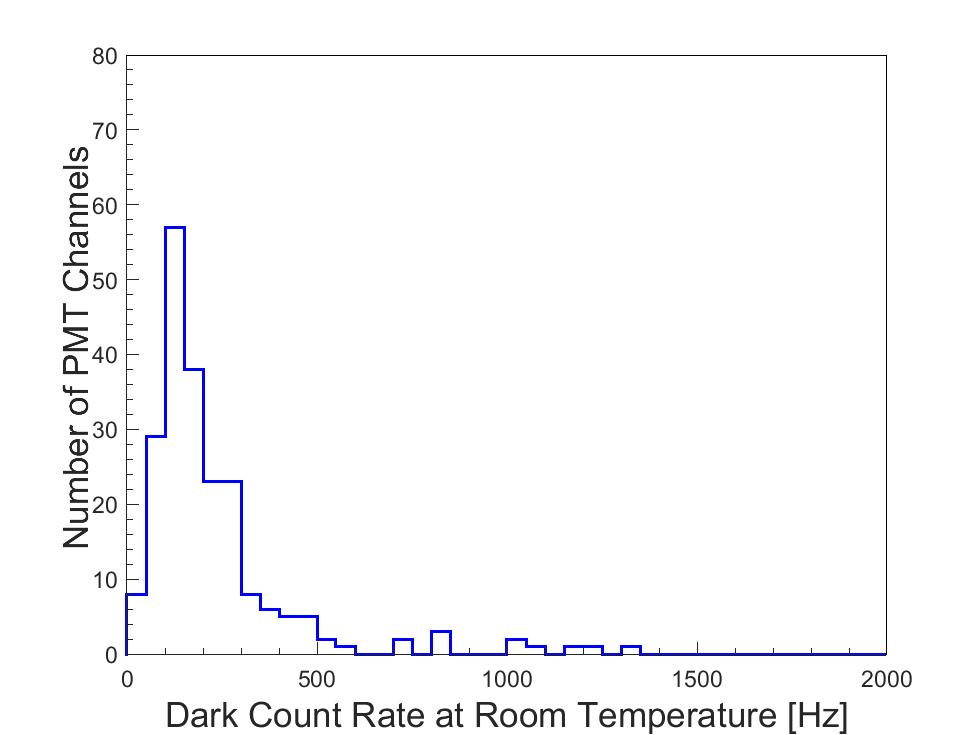}
        \qquad
        \includegraphics[width=0.45\textwidth]{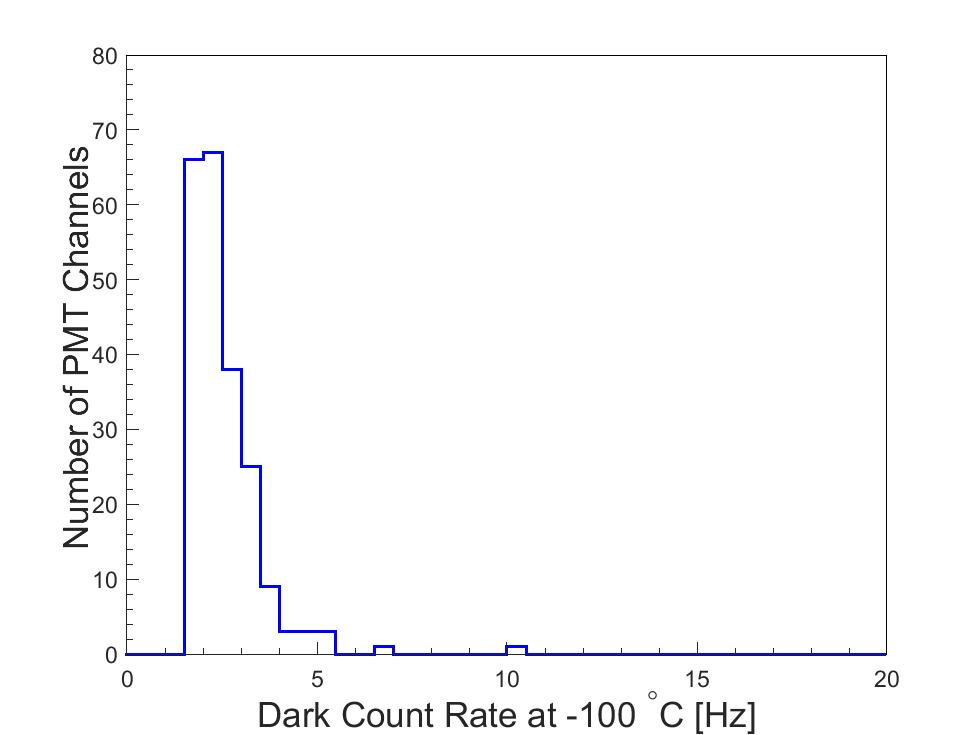}
        \caption{(Left) Distribution of the DCRs of 216 channels of the 54 tested R12699 PMTs at -1000~V in room temperature. (Right) DCRs of the same PMTs at~-100$^{\circ}$C with the same HV value.}
        \label{fig:DC_distribute} 
    \end{figure}

    Prior to the test, all PMTs under qualification are subjected to a 24-hour period in a dark environment to mitigate the residual effect of previous photon exposure. The dark count threshold is carefully set at 2.5~mV to effectively differentiate dark signals from background noise. Fig.~\ref{fig:DC_distribute} presents the histograms of the DCRs measured for all tested R12699 PMT channels at a bias of -1000~V in both room temperature and -100~$^{\circ}$C. The mean value of the DCR at room temperature for all tested R12699 PMTs is 229.6~Hz per channel. After cooling to -100~$^{\circ}$C, the average DCR is significantly reduced to 2.5~Hz per channel.
	
    \subsection{After-pulse probability}
    \label{subsec:APP}
	
    AP is a phenomenon typically characterized by PMT signals that occur after an initial PE signal within a defined time range. APs result from positive ionized residual gas molecules striking the cathode, a phenomenon known as ion feedback \cite{f}. A high AP probability suggests PMT leakage or a large amount of residual gas. Occasionally, AP signals may also include some dark signals and background noise. In specific time delays, such as the range of 0.2 to 5~$\rm \mu$s, AP signals can mimic the detection of $S2$ signals, thereby contributing to additional background noise or fluctuating the deposit energy reconstruction of $S2$ signals. Therefore, a low AP probability is imperative for the precision of $S2$ signal analysis.
    
    \begin{figure}[htbp]
        \centering
        \includegraphics[width=0.7\textwidth]{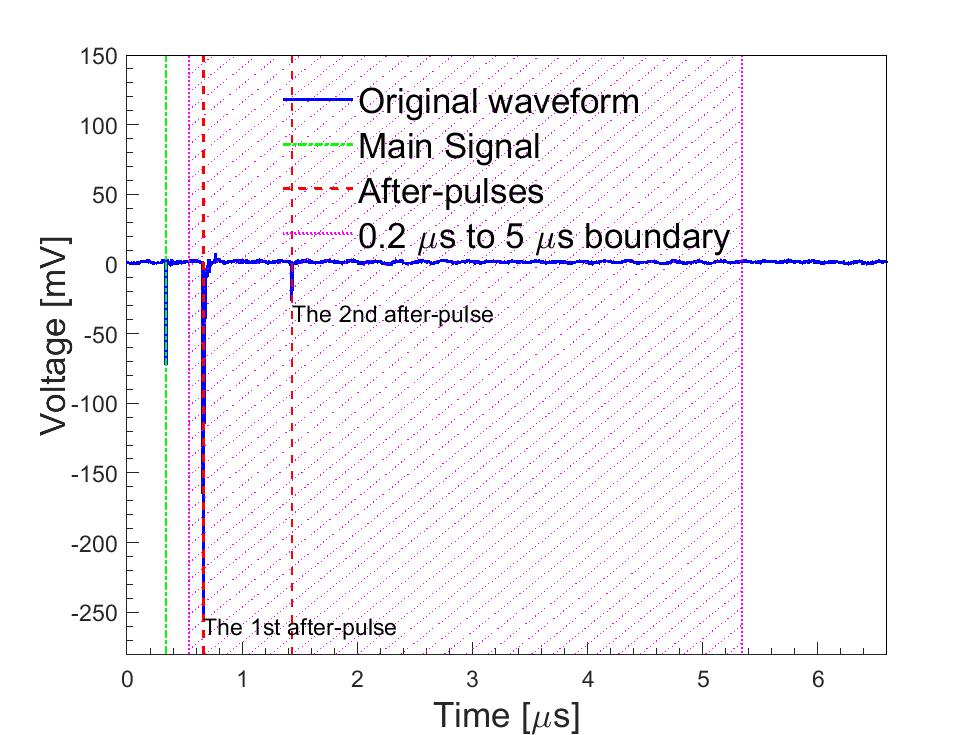}
        \caption{A typical AP signal waveform. The main signal at 0.34~$\mu$s is indicated by green dash dot lines, and two AP signals appear at 0.33~$\mu$s and 1.09~$\mu$s after the main signal.}
        \label{fig:AP_distribute} 
    \end{figure}
    
    In the assessment of the AP probability, the FADC is activated by SPE signals, with the data acquisition window extends to 8~$\mu$s post-trigger. Fig.~\ref{fig:AP_distribute} illustrates a typical AP waveform containing 2 APs. The AP events are quantified by enumerating the PE signals observed within the 0.2 to 5~$\mu$s interval subsequent to the primary signal and adjusted by subtracting the anticipated number of dark signals, which is derived from the DCR and the duration of the observation windows. The findings indicate that all 54 units of R12699 PMTs have met the AP probability criteria, with the average AP probability recorded across the 216 channels is 0.5\%.

    \section{Summary and discussion}
    \label{sec:sum}
    
    In this article, we detail the background screening and the electrical performance in the room temperature and the cryogenic temperature of R12699 PMTs. By the replacement of two high radioactive materials, Kovar and Glass-1, the PMT achieves extremely low levels of $\rm^{60}Co$ (\textasciitilde 0.08 mBq/PMT), $\rm^{232}Th(l)$, $\rm^{238}U(l)$ (\textasciitilde 0.06 mBq/PMT), along with other radioisotopes. The radon emanation rate of the R12699 PMT is $<$3~$\rm \mu Bq/PMT$ and its surface radioactivity is $<$18.4~$\rm \mu Bq/cm^{2}$. The $^{40}$K radioactivity in the R12699 PMT remains high as shown in Tab.~\ref{tab:pmt_comparsion}, and we plan to improve it in the future. $^{40}$K comes from the material of photocathode, which contains potassium. It has been demonstrated that $^{39}$K-enriched potassium chromate can be used to produce the photocathode which can significantly reduce $^{40}$K activity~\cite{XMASS}. Therefore, we plan to replace the current photocathode material to further lower the radioactivity of the R12699 PMT.

    As for the evaluation with cryogenic test, the 216 individual channels of 54 tested PMTs exhibit an average gain of $(4.23\pm 1.37)\times 10^6$ at -100~$^{\circ}$C and $(4.18\pm 1.27)\times 10^6$ at room temperature, indicating a negligible difference. These PMTs demonstrate a low average DCR of $(2.5\pm 0.9)$~Hz per channel and the average AP probability of 0.5\% at -100~$^{\circ}$C.

    \begin{table}[htbp]
        \caption{The mean values or operated values of characteristics of R11410 and R12699 PMTs at -100~$^{\circ}$C. The values of R11410s are from~\cite{i,j,n,u}.} 
        \label{tab:11410_vs_12699}
        \centering
        \resizebox{\textwidth}{!}{
        \begin{tabular}{ccccc}
            \hline
            \multirow{2}{*}{PMT model} &   \multicolumn{4}{c}{Characteristics}   \\
            \cline{2-5}
                & Gain  & AP probability &  DCR/channel [Hz] & DCR/area [Hz/cm$^2$]    \\
            \hline
            R11410-21 (XENON1T)       & $5.8\times 10^6$ at 1500~V & 1.4\% & 44 & 0.97\\
            R11410-20/22 (LZ)          & Operated at $3.5\times 10^6$ & - & - & - \\
            R11410-23 (PandaX-4T)      & Operated at $5\times 10^6$ & 1\% & 20 & 0.44\\
            R12699-406-M4      & $4.23\times 10^6$ at -1000~V & 0.5\% & 2.5 & 0.32\\
            \hline
        \end{tabular}
        }
    \end{table}

    Tab.~\ref{tab:11410_vs_12699} shows the characteristics of the R11410 PMTs measured by XENON1T \cite{j}, LZ \cite{n,u} and PandaX-4T \cite{i}, alongside the results for R12699 PMTs. The comparison reveals that the model of R12699 exhibits a nearly equivalent gain and AP probability to the R11410 model, while the DCR per channel and DCR per unit area at -100~$^{\circ}$C of the R12699 PMTs are lower than those measured for the R11410. The features with low radioactivity and robust electrical performance in cryogenic temperature make the R12699 PMTs ideal for next-generation liquid xenon detectors and other rare event searches. 

\section*{Acknowledgement}
This work has been supported by the Ministry of Science and Technology of China (No. 2023YFA1606203 and 2023YFA1606204), the National Natural Science Foundation of China (No. 12222505), Shanghai Pilot Program for Basic Research-Shanghai Jiao Tong University (No. 21TQ1400218), National Science Foundation of Sichuan Province (No. 2024NSFC1371). We also thank the sponsorship from the Chinese Academy of Sciences Center for Excellence in Particle Physics (CCEPP), Hongwen Foundation in Hong Kong, New Cornerstone Science Foundation, Tencent Foundation in China, and Yangyang Development Fund. We appreciate Hamamatsu Photonics K.K. for their dedicated efforts and detailed work in developing this novel low-background PMT.

\bibliography{mybibfile}
\end{document}